\documentclass[12pt,nofootinbib,preprint,aps]{revtex4}

\usepackage{epsfig}

\def  \bcen     {\begin{center}}
\def  \ecen     {\end{center}}
\def  \beq      {\begin{equation}}
\def  \eeq      {\end{equation}}
\def  \beqa     {\begin{eqnarray}}
\def  \eeqa     {\end{eqnarray}}

\def  \bfleft   {\begin{flushleft} }
\def  \efleft   {\end{flushleft} }
\def  \bfright  {\begin{flushright} }
\def  \efright  {\end{flushright} }

\def  \pptotlnu {p p \to \ell_H \nu_H}
\def  \pptotll  {p p \to \ell_H \bar{\ell}_H}
\def  \et       {\not\!\! E_T}

\begin{document}


\title{T-parity odd heavy leptons at LHC}
\author{Giacomo Cacciapaglia  \footnote{g.cacciapaglia@ipnl.in2p3.fr}}
\affiliation{Universit\'e de Lyon, France; Universit\'e Lyon 1,
  CNRS/IN2P3, UMR5822 IPNL, F-69622 Villeurbanne Cedex, France,}
\author{S. Rai Choudhury \footnote{srai.choudhury@gmail.com}}
\affiliation{Center for Theoretical Physics (CTP), Jamia Millia University,
Delhi, \& \\
Indian Institute of Science Education \& Research (IISER),
Govindpura, Bhopal, India.}
\author{Aldo Deandrea \footnote{deandrea@ipnl.in2p3.fr}}
\affiliation{Universit\'e de Lyon, France; Universit\'e Lyon 1,
  CNRS/IN2P3, UMR5822 IPNL, F-69622 Villeurbanne Cedex, France,}
\author{Naveen Gaur \footnote{gaur.nav@gmail.com}}
\affiliation{Department of Physics \& Astrophysics, University of Delhi,
  Delhi - 110007, India.}

\begin{abstract}
\noindent  Little Higgs models with T-parity can easily satisfy electroweak 
precision tests and at the same time give a stable particle which is a candidate for cold dark matter.
In addition to little Higgs heavy gauge bosons, this type of models predicts a set of new T-odd fermions, 
which may show quite interesting signatures at colliders. 
We study purely leptonic signatures of T-odd leptons at the Large Hadron Collider (LHC).
\end{abstract}

\maketitle



\section{Introduction \label{section:1}}

The Standard Model (SM), when put in a more general context, is affected by  
the hierarchy problem, since within the SM the Higgs boson gets a
quadratically divergent contribution to its mass if the model is
considered as an effective theory only valid up to some high energy
scale. If one considers a situation in which physics is perturbative,
precision electroweak physics indicates that the Higgs mass cannot be
very large and that the effective cutoff is preferably heavier than $5-10$ TeV~\cite{littlehierarchy}. 
This requires a fine tuning from the eventual scale of new physics to the
electroweak scale. Little Higgs models are effective theories based on
the non-linear sigma model structure where the Higgs field is a
Nambu-Goldstone Bosons (NGB) of a global symmetry which is
spontaneously broken at some higher scale by an expectation value $f$
(for a review, see \cite{lhrev} and references therein). The Higgs
field acquires its mass through symmetry breaking at the electroweak scale
and, protected by the approximate global symmetry, it remains
light. The original Little Higgs models, however, are disfavoured by electroweak
precision tests \cite{precisionEW}, which push the scale $f$ above few
TeV thus restoring the fine tuning problem.  
On the other hand, models with an extra parity
(called T-parity \cite{tpar}) are in agreement with present
constraints while allowing a scale $f$ which is sufficiently light to
be in the LHC range and solve the little hierarchy problem~\cite{tparprec}. T-parity
is also motivated by the fact that it offers a stable candidate for dark
matter~\cite{tparDM}. Generic little Higgs theories predict, at a scale of the order of 1
TeV, new particles responsible for cancelling 
the standard model quadratic divergences at 1-loop: heavy weak gauge
bosons, new heavy scalars, new fermions that are partners of the top
quark and partners of the light fermions. These new particles are  
charged under the T-parity. In addition there are also T-odd doublet
partners for every standard  
model fermionic doublets. As a typical example we consider the 
Littlest Higgs model with T-parity (LHT). 
In this work, we study the possibility of observing the T-odd
heavy leptons in the single ($p p \to \ell_H \bar{\nu}_H$) and pair
production ($p p \to \ell_H \bar{\ell}_H$) of heavy charged T-odd
leptons at LHC.  
In particular, we will focus on the purely leptonic decay modes.
The ($p p \to \nu_H \bar{\nu}_H$) production, even though it may give rise to 
a four charged lepton channel, is less interesting because the smaller 
cross section, together with small leptonic branching ratios, renders 
the signal too feeble at the LHC.

Our paper is organised as follows: in section \ref{section:2} we
discuss the main features of the model we have considered for our
analysis and also our framework for
event generation and detector simulations. 
In section \ref{section:3} we compute the production of
single charged T-odd lepton in the channel $p p \to \ell_H
\bar{\nu}_H$. 
In section \ref{section:4} we discuss pair production of
T-odd charged leptons.
In section \ref{section:5} we discuss the pair production of
T-odd neutral leptons. Finally we conclude in
section \ref{section:6} with the summary of the results.


\section{T-odd leptons in LHT \label{section:2}}

In the following, we briefly review the main features of the Littlest
Higgs model with T-parity and in particular  of
the T-odd heavy fermions. The model is based on a $SU(5)$ global symmetry, of which a
$[SU(2)_1\times U(1)_1] \times [SU(2)_2\times U(1)_2] $ subgroup is gauged.  
A discrete parity (T-parity) exchanges the two $[SU(2) \times U(1)]$
groups. At the scale $f$, the global symmetry is spontaneously broken
down to a $SO(5)$ group resulting in 14 massless Nambu-Goldstone (NG)
bosons and the gauged symmetry is reduced to its diagonal $SU(2)_L\times U(1)_{\rm Y} $ 
subgroup identified with the standard model gauge group. 
The lightest heavy gauge boson, $A_H$, is the partner of the photon and
is generally the lightest stable T-odd particle in the model.

The implementation of T-parity in the fermion sector requires that
each standard model fermion doublet is replaced by the 
fields $F_i\ (i =1,2)$ \cite{tpar,Hubisz:2004ft}, where each 
$F_i$ is a doublet under one $SU(2)_i$ and a singlet under the other. T-parity
simply exchanges $F_1 $ and $F_2$. The T-even combination of $F_i$ is identified with
the standard model fermion doublet while the other (T-odd) one
is its heavy partner $(F_H)$. Mass terms for these T-odd
heavy fermions are generated by Yukawa interactions with additional T-odd 
$SU(2)$ singlet fermions. Assuming a universal and flavour diagonal
Yukawa coupling $\kappa_\ell$, for $l_H $ and $\nu_H$ (the
T-odd heavy partners of the standard model leptons), we have the following
masses 
\beq
m_{l_H} = \sqrt{2} \kappa_l f\ ,  \quad m_{\nu_H} = \sqrt{2} \kappa_l f \left(1 - \frac{v^2}{8 f^2}\right)\ ;
\eeq
as the scale $f$ is typically of the order 500 GeV or larger, 
it is clear that the T-odd heavy partners have nearly equal masses
as they are only split by $v^2/f^2$ effects.  
With the simplifying assumption of universal and flavour diagonal
Yukawa couplings we therefore have only two free  
parameters: the new mass scale $f$ and the flavour independent Yukawa
coupling $\kappa_\ell$. 
We use in the following the Feynman rules of mirror fermions in
accordance with \cite{Goto:2008fj}.  
These modifications in the couplings of T-odd fermions provide the correct result  
at the order $v^2/f^2$ for the cancellation of the divergences in
Z-penguin diagrams in various flavour changing  
decays. The Yukawa coupling $\kappa_\ell$ in general depends on
flavour and this can in turn generate Lepton Flavour Violation (LFV)
in this class of models  
\cite{Choudhury:2006sq}. For our analysis we will
assume that $\kappa_\ell$ is flavour blind and universal, hence it
does not give rise to new sources of flavour violation.


\subsection{ Calculation and event generation details  }

The cross-sections and branching ratios for T-odd lepton production
and decays have been calculated with CalcHEP v2.5.4
\cite{Pukhov:2004ca}. For this purpose we have used the modified LHT
model file provided in \cite{Belyaev:2006jh}. In the modified model file
we included the changes in the Feynman rules of mirror fermions in
accordance with \cite{Goto:2008fj} \footnote{The revised
LHT model files can be obtained from 
{\tt {http://deandrea.home.cern.ch/deandrea/LHTmodl.tgz}}.}.  The LHC
cross-sections were calculated for the center of mass energy of
10 and 14 TeV. We have used CTEQ6L PDFs (parton distribution
functions) with QCD coupling scale set to $\sqrt{\hat{s}}$. 

The event simulations always refer to the 14 TeV case: the signal 
events were generated using Calchep v2.5.4 and were
interfaced to {\sc pythia} 6.4.21 \cite{Sjostrand:2006za} by Les
Houches Event interface (LHE) \cite{Alwall:2007mw}. The ISR/FSR
switches in {\sc pythia} were kept on in the simulations. In order to
make more realistic estimates we have further passed the events through
the fast ATLAS detector simulator {\sc atlfast} \cite{atlfast} for
realistic detector effects. {\sc atlfast} identifies isolated leptons,
b and $\tau$ jets. It also reconstructs missing energy. The jets in
{\sc atlfast} are reconstructed using a simple cone algorithm. The SM 
backgrounds, on- and off-shell W production and $ZZ$ production,
were generated using {\sc pythia} 6.4.21 and were further processed
through {\sc atlfast}. The $WWW$ events were generated using {\sc
  madgraph} \cite{Maltoni:2002qb} and were interfaced to {\sc pythia}
6.4.21 for ISR/FSR effects via LHE. The events thus generated were
further passed through {\sc atlfast}.

\subsection{Decay of T-odd leptons}

\begin{figure}[htb]
\begin{center}
\hspace{-2cm}
\epsfig{file=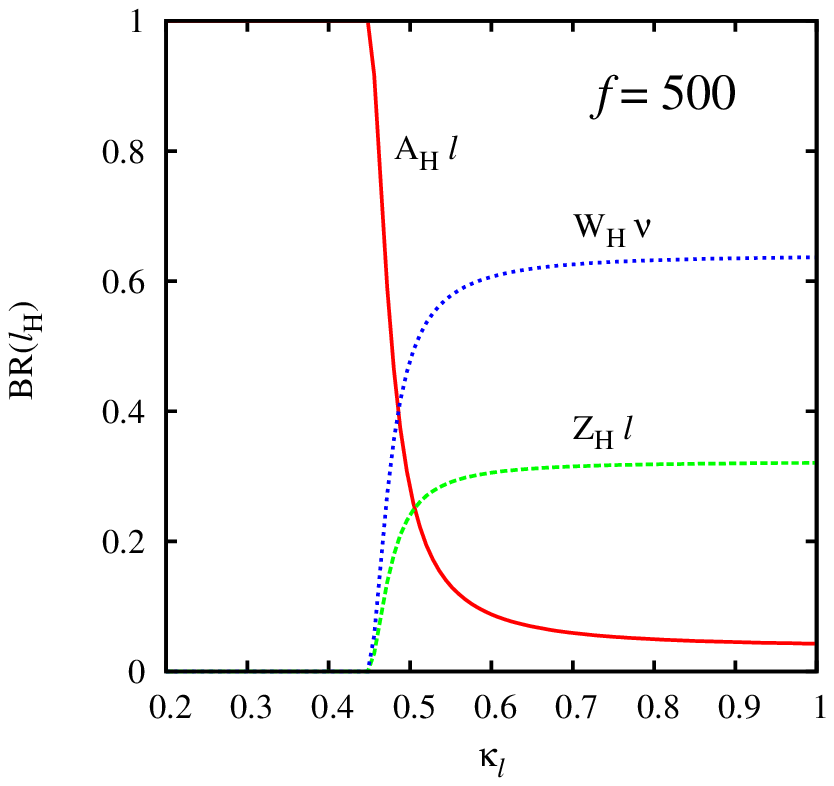,width=.63\textwidth}
\hspace{-3cm}
\epsfig{file=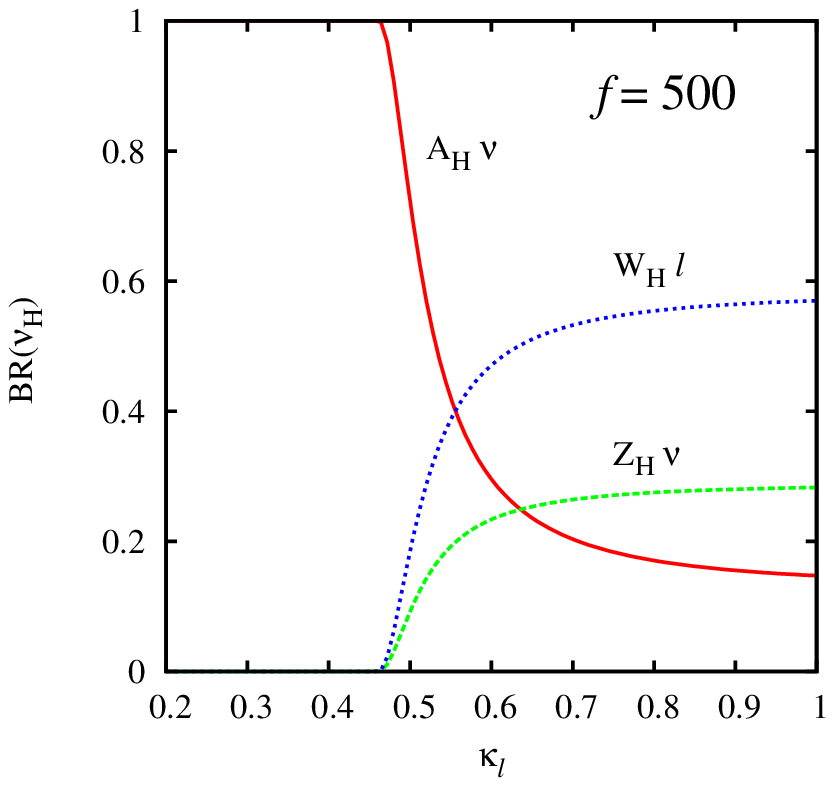,width=.63\textwidth}
\caption{\sl Branching ratios of $\ell_H$ (left) and $\nu_H$ (right) as
  a function of $\kappa_\ell$ for fixed $f$. As can be seen from these
figures the branching ratios in channel $(\ell_H, \nu_H) \to (\nu,
 \ell) W_H$ grow until $\kappa_\ell = 0.55$ (note this decay
mode is responsible for trilepton signal) and then flatten out. The
above branching fractions remain the same for nearly all values of $f$. }
\label{fig:br31}
\end{center}
\end{figure}

The input parameters of the LHT model that are relevant for our analysis
are the scale $f$ and the coupling $\kappa_\ell$. 
As can be seen from Figure \ref{fig:br31}, in the region $\kappa_\ell < 0.46$,
the branching ratios are:
\beq
BR(\ell_H \to A_H \ell) = 1\ , \quad
BR(\nu_H \to A_H \nu) = 1\ .
\eeq
For $\kappa_\ell > 0.46$, the leptons became heavier than the gauge bosons $W_H$ and $Z_H$ and other modes start opening up: when $\kappa_\ell \geq 0.5$, the dominant decay modes become $\ell_H \to W_H \nu$ and
$\nu_H \to W_H \ell$. It is known that in this range
($\kappa_\ell \geq 0.5$) $W_H$ decays to $W A_H$ with almost 100\%
branching ratio \cite{Datta:2007xy,Cao:2007pv}.
 
\begin{figure}[htb]
\begin{center}
\hspace*{-1cm}
\epsfig{file=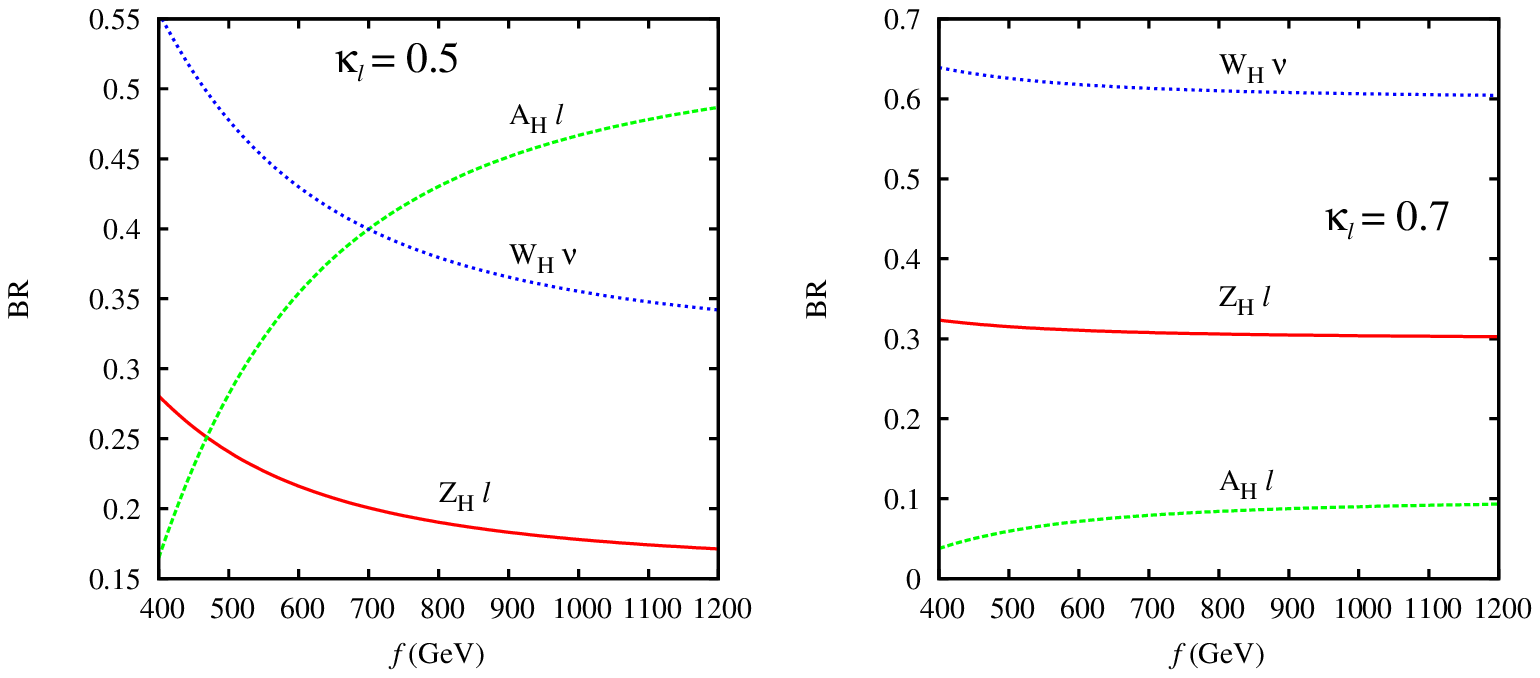,width=1.1\textwidth}
\vskip -.3cm
\caption{\sl Branching ratios of $\ell_H$ decay as a function of $f$.} 
\label{fig:br1}
\vspace{1cm}
\hspace*{-1cm}
\epsfig{file=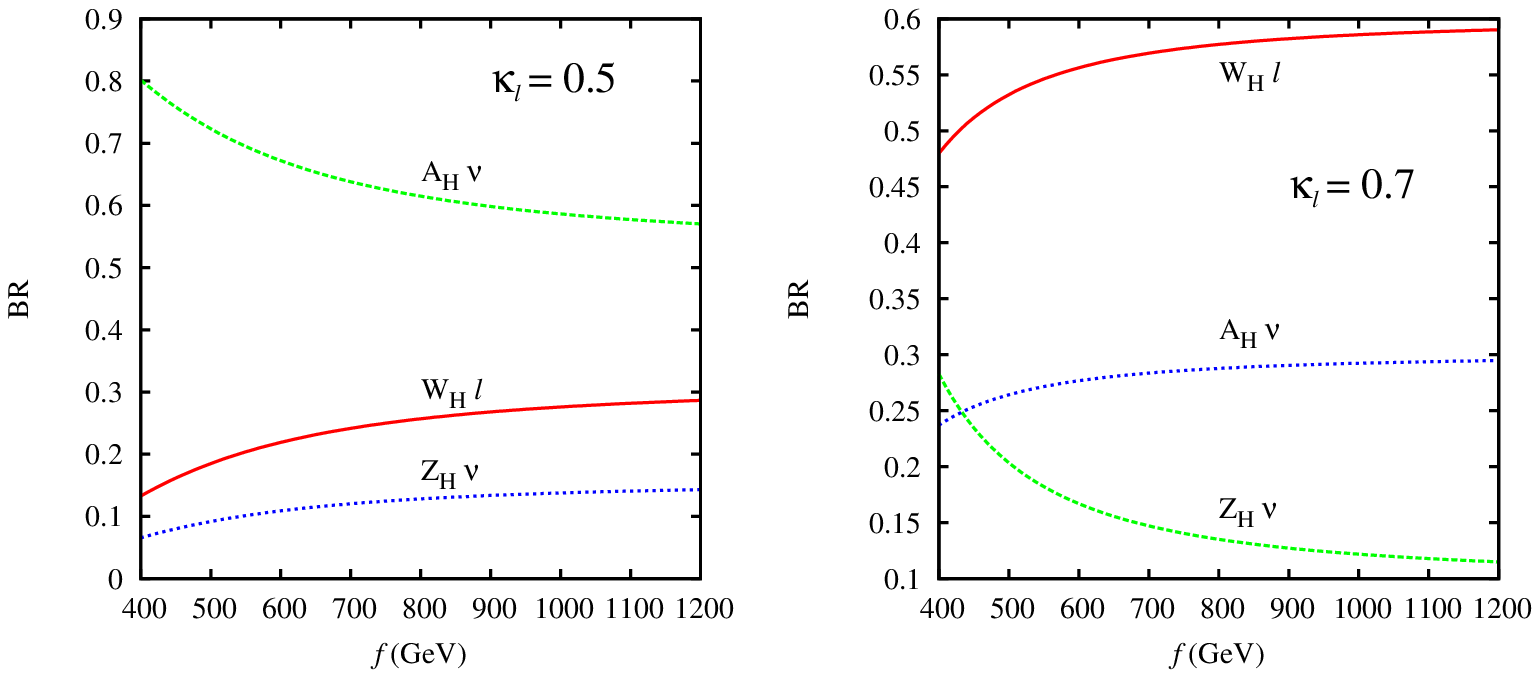,width=1.1\textwidth}
\vskip -.3cm
\caption{\sl Branching ratios of $\nu_H$ decay as a function of $f$.} 
\label{fig:br2}
\end{center}
\end{figure}

We show the dependence of the T-odd lepton branching ratios on
the symmetry breaking scale $f$ in Figure \ref{fig:br1} for charged
T-odd lepton ($\ell_H$) and in Figure \ref{fig:br2} for neutral
T-odd lepton ($\nu_H$) for some indicative values of
$\kappa_\ell$. 
Considering the SM $W$
decaying leptonically, the dominant decay chains in
the region $\kappa_\ell \geq 0.5$ (with a combined probability of about 12\% each) are:
\beqa
\ell_H^\pm &\to& W_H^\pm (\to W^\pm A_H) \nu \to \ell^\pm \et \ , \nonumber \\
\nu_H^\pm &\to& W_H^\pm (\to W^\pm A_H) \ell^\mp \to \ell^\pm \ell^\mp \et  \ .
\label{eq:decay-chain}
\eeqa
Note that we are considering the decay chains with only
leptons\footnote{By leptons in our analysis we mean electrons and
  muons.} in the final state, the reason being that with purely leptonic final state
it is easier to suppress the backgrounds coming from $t
\bar{t}$ and other QCD processes by imposing jet veto on the events.   
Armed with the information about the possible decay channels of
T-odd leptons, in next sections we will discuss the 
production of these heavy leptons at LHC. 

\section{Single charged T-odd lepton production ($p p \to \ell_H
  \bar{\nu}_H$) \label{section:3}}   

\begin{figure}[htb]
\begin{center}
\epsfig{file=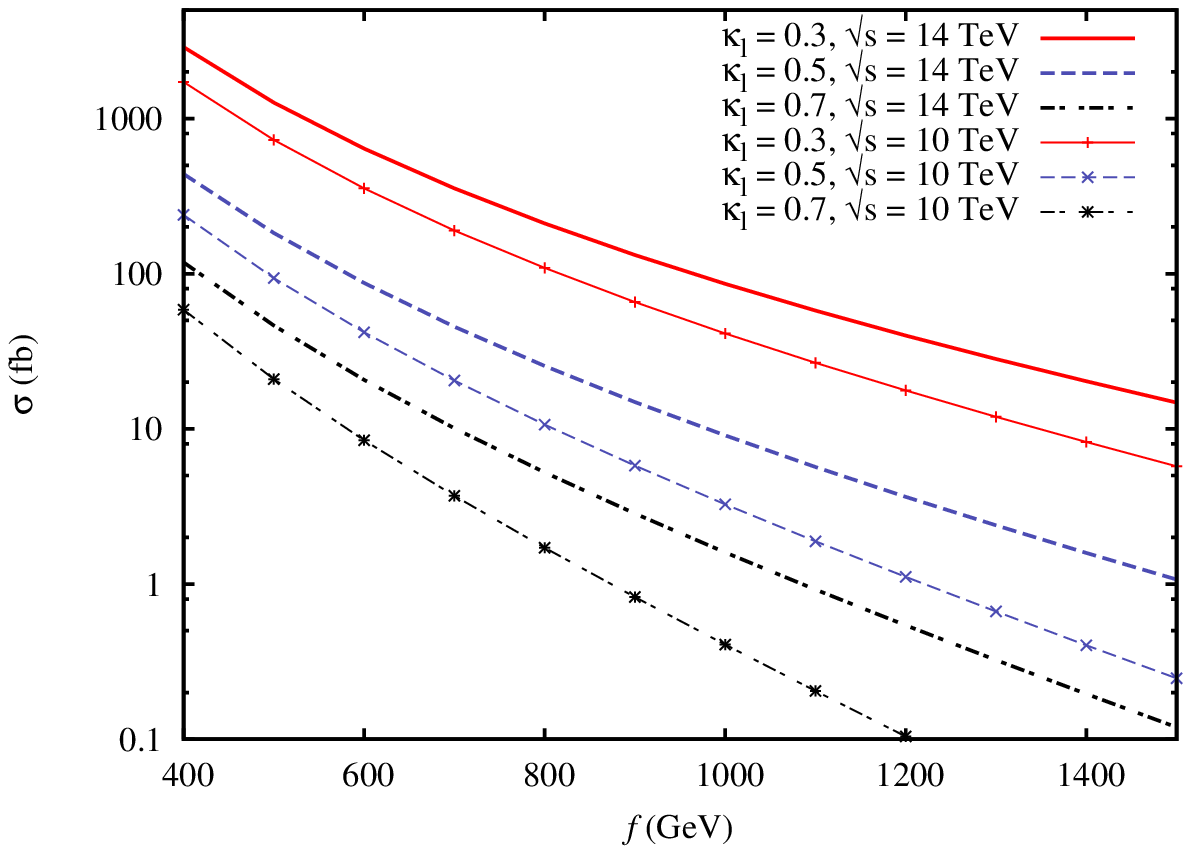,width=.8\textwidth}
\caption{\sl LHC production cross-sections for the process $\pptotlnu$ for the options of centre of mass energy of 14 TeV and 10 TeV, and for different benchmark points as a function of $f$.}
\label{fig:cs1}
\end{center}
\end{figure}

In a reasonable range of parameters of the LHT model, it is possible to
produce a single charged T-odd lepton in association with the
T-odd heavy neutrino at LHC. 
Initially LHC is expected to run at low energy $\sqrt{s} = 7$ TeV where
it is expected to collect a small luminosity ${\cal L} < 100$ pb$^{-1}$. This
energy will be upgraded to $\sqrt{s} = 10$ TeV with an expected integrated
luminosity of the order ${\cal L} = {\cal O} (100)$ pb$^{-1}$, before reaching the design energy of 14 TeV. 
In Figure \ref{fig:cs1} we have shown the production cross-sections of
$p p \to \ell_H \bar{\nu}_H$ for the center of mass energy of 10 and
14 TeV. Depending on the energy, it can exceed a picobarn in part of
the $(f,\kappa_\ell)$  parameter space. 

The decay chains depend on the parameter region: as discussed 
in the previous section, in the region $\kappa_\ell < 0.46$
the T-odd leptons decay directly in the heavy photon, $\ell_H \to A_H \ell$ and
$\nu_H \to A_H \nu$. For $\kappa_\ell \geq 0.5$ the T-odd
leptons decay mainly via heavy charged $W_H$' s, as in eqns (\ref{eq:decay-chain}). 
These two regions of parameter space give different signatures,
therefore we will discuss them separately in next sub-sections. 


\subsection{$\kappa_\ell < 0.46$} 

In this region the T-odd leptons decay with 100\% branching
ratio to the heavy photon and the corresponding SM lepton.
The heavy neutrinos, therefore, decay invisibly: this gives 
rise to the signature of a single isolated charged lepton
with missing energy ($\ell^\pm \et$). 
The dominant SM background for this signature comes from the on- and
off-shell $W$ production:
\begin{itemize}
\item{} on-shell $W$ production, $p p \to W^\pm \to \ell^\pm
  \bar{\nu}$. 
\item{} off--shell $W$ production  $p p \to W^\pm Z$ with the $W$ going
  to leptons and the $Z$ decaying invisibly via $Z \to \nu \bar{\nu}$. 
\end{itemize}
This signature has been already analysed in \cite{Godbole:2003it,
  Cao:2004tu}. They argued that it may be possible to 
reduce the backgrounds by using high $\et$ cuts and by using the
transverse mass cuts. The transverse
mass is defined as follows:
\beq
m_T = \sqrt{2 p_T^\ell E_T^{miss} (1 - cos(\phi))} \ ,
\label{eq:mt1}
\eeq
where $\phi$ is the angle between the transverse momentum of the lepton $p_T^\ell$ and the transverse
component of missing energy. 

\begin{figure}[htb]
\begin{center}
\epsfig{file=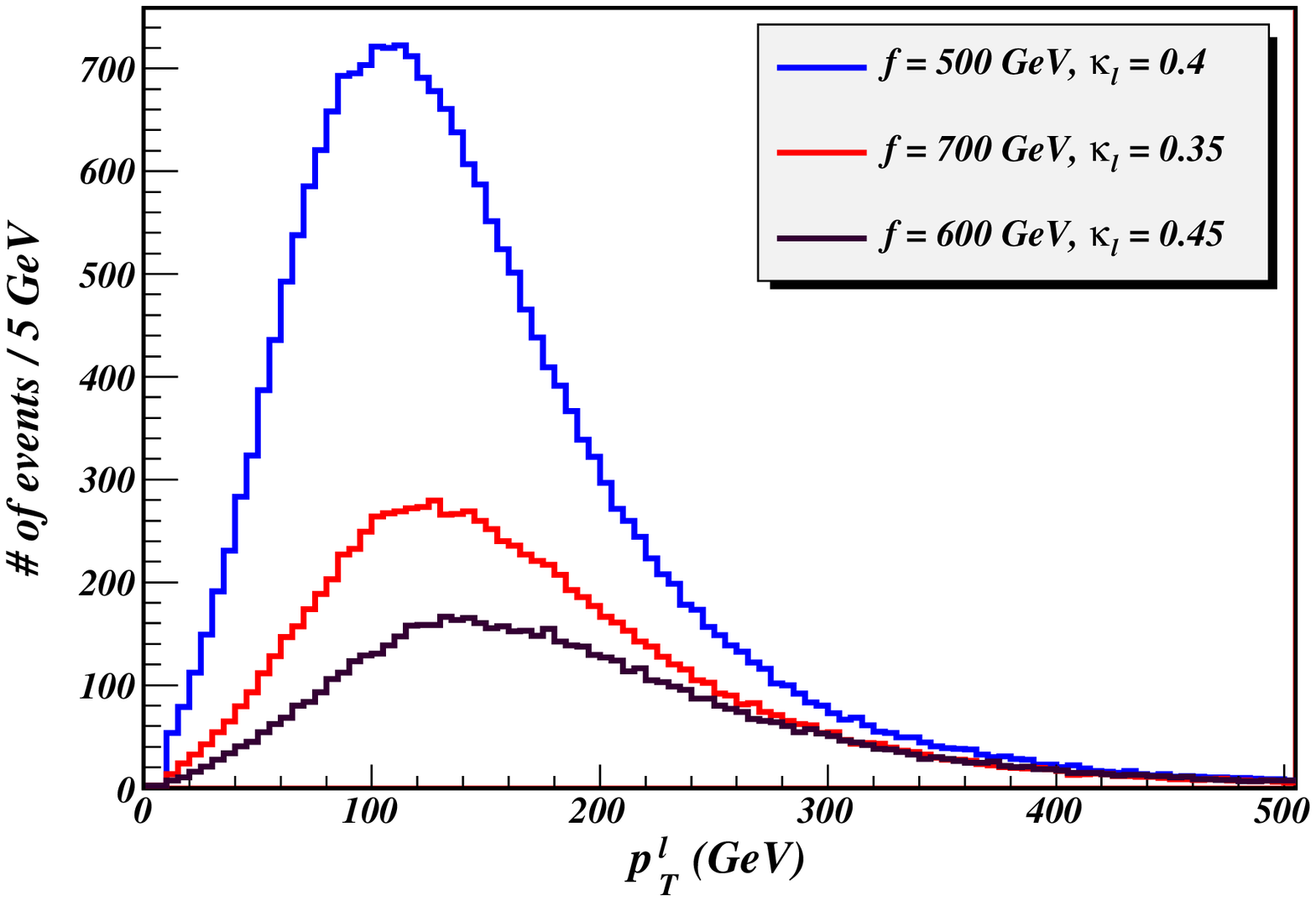,width=.52\textwidth} 
\hspace{ -.7cm}
\epsfig{file=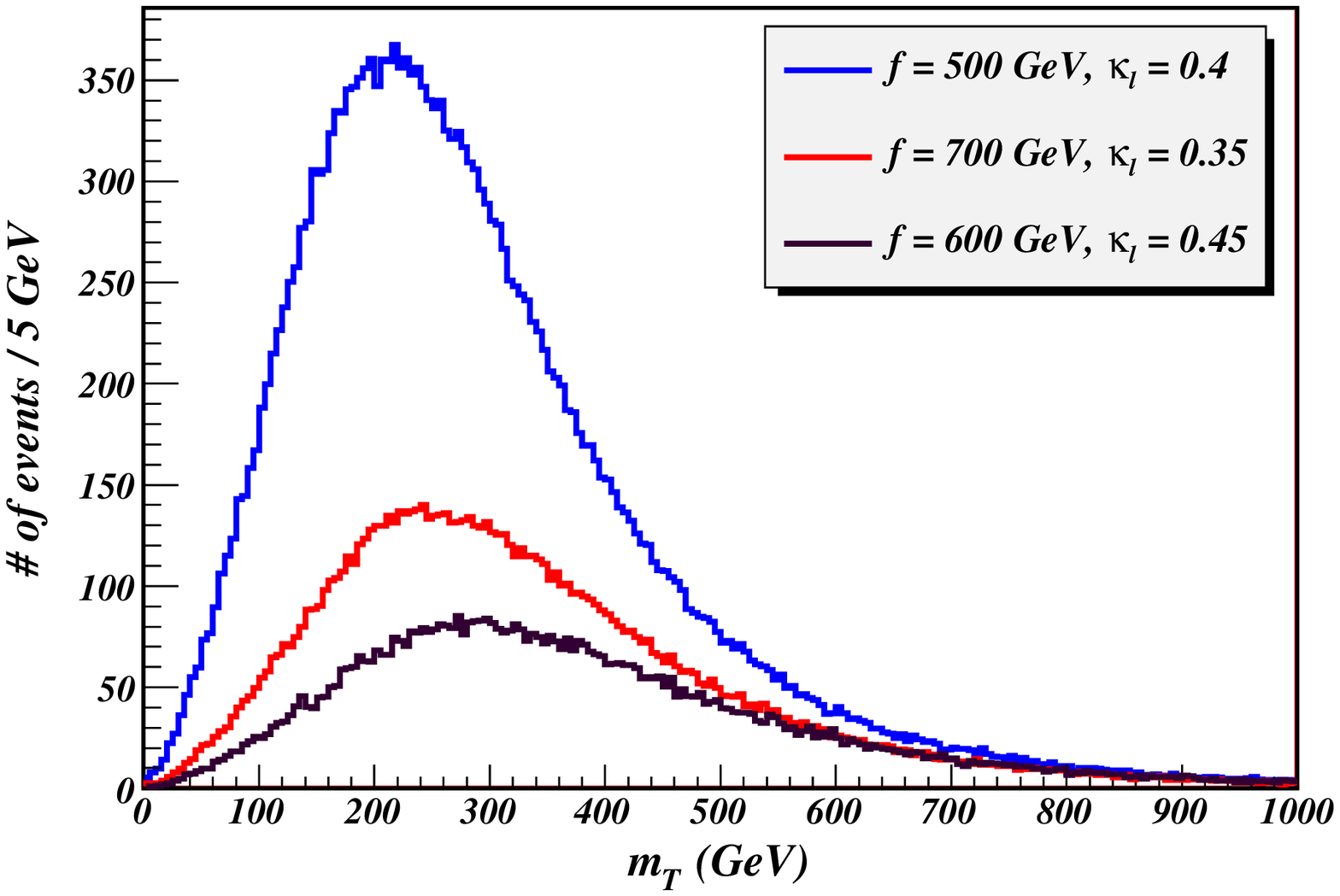,width=.5\textwidth}
\caption{\sl $p_T^\ell$ distribution (left) and $m_T$ distribution
  (right) for LHT. The model parameters are given in legends.
  In these distributions we have taken the LHC luminosity to be ${\cal
  L} = 100$ fb$^{-1}$. }
\label{fig:tlep1}
\end{center}
\end{figure}

\noindent In order to perform the analysis we impose the following
{\sl pre-selection cuts}:
\begin{itemize}
\item[(a)] jet veto: we reject events having any
  resolved jet. By resolved jet we mean a jet that is visible in
  the detector. For this we veto an event having a jet with $p_T > 30$ GeV
  and rapidity $|\eta| < 3$. 
\item[(b)] exactly one charged lepton with $p_T^\ell > 10$ GeV and
  $|\eta| < 3$.
\end{itemize}

\noindent The $p_T^\ell$ and $m_T$ distributions after the {\sl preselection cuts} 
for signal events are given in Fig \ref{fig:tlep1}.

\begin{table}[htb]
\begin{center} 
\begin{tabular}{|c | c | c || c | c | c | c | c |} 
\hline 
Parameters $\Rightarrow$ & SM &  SM  & $f = 700$ &  $f= 600$ & $f = 500$ & $f = 600$ & $f = 700$ \\ 
Cuts $\Downarrow$                     & on shell & off shell & $\kappa_\ell = 0.4$ & $\kappa_\ell =
                               0.4$ & $\kappa_\ell = 0.4$ 
                               & $\kappa_\ell = 0.45$ & $\kappa_\ell = 0.35$\\ \hline 
$\sigma$ (fb)  &  &  &  114.8   &
212.8    &  433  & 133.55 & 195.16  \\ \hline 
Presel. cuts & 1.7 $\times 10^9$ & 6.1 $\times 10^4$ & 5823.1 &  11067.9  & 22882.8 & 6800.5 & 10072.7 \\ \hline
$p_T^\ell > 100$ GeV & 4.9$\times 10^5$ &  2137.4 & 4916.4 & 8656.8 & 15652.2 & 5687.9 & 7852.9 \\
$m_T >$ 200 & 3.16 $\times 10^5$ & 1818 & 4849.8 & 8481.5 & 15212.2 & 5604.8 & 7689 \\ 
$m_T >$ 300 & 8.17 $\times 10^4$  & 451 & 3623.8 & 5722.9 & 8887.6 & 4138.4 & 5157.2 \\ 
$m_T >$ 400 & 2.72 $\times 10^4$  & 147 & 2343.7 & 3335 & 4593 & 2664.7 & 3001.2  \\ \hline 
$p_T^\ell > 200$ GeV & 2.82$\times 10^4$ & 147.9 & 2350.1 & 3351.6 & 4607.1 & 2673.1 & 3011.3 \\
$m_T >$ 300 & 2.82$\times 10^4$ & 146.8 & 2349.2 & 3350.3 & 4604.5 & 2672.2 & 3009.2 \\ 
$m_T >$ 400 & 2.72$\times 10^4$ & 139.4 & 2278.7 & 3238.2 & 4422 & 2592 & 2904.7  \\ \hline 
$p_T^\ell > 300$ GeV & 3026.7 & 26.8 & 866.2 & 1074.2 & 1284.3 & 966.5 & 949.6 \\
$m_T >$ 400 & 30.3 & 0.26 & 8.7 & 10.7 & 12.8 & 9.7 & 9.5 \\ \hline 
$S/\sqrt{B}$ (10 fb$^{-1}$) & - & - & 5 &  6.2 & 7.3 & 5.5 & 5.4 \\ 
$S/\sqrt{B}$ (100 fb$^{-1}$) & - & - & 15.6 &  19.5 & 23.2 & 17.5 & 17.2 \\ 
\hline 
\end{tabular}
\caption{\sl Results of the simulations for the signal $\ell^\pm \et$. The
  numbers of events, after sequentially imposing
  the cuts mentioned in the text, are for an LHC luminosity ${\cal L} = 100$ fb$^{-1}$. 
The most efficient cut is $p_T^\ell > 300$ GeV, for which we also give $S/\sqrt{B}$ for different LHC luminosities.
 Note that a cut at the same level on $m_T$
is practically ineffective as the two quantities are strongly
correlated. Harder cuts, for example $m_T > 400$ GeV, start
affecting more the signal events, therefore reducing $S/\sqrt{B}$.} 
\label{table:1}
\end{center}
\end{table}

\noindent As can be seen from the results given in Table \ref{table:1}
the backgrounds are huge as compared to signal. To extract signal from
such huge backgrounds, we try to impose additional cuts. In the case of
the LHT model, the charged lepton comes from the decay of heavy T-odd
lepton and hence would have a relatively high $p_T$ as compared to SM
leptons. 
Hence, to further reduce the SM backgrounds we have used the following
secondary cuts: 
\begin{itemize}
\item{} $p_T^\ell$ cut:  we imposed three different
  values, namely 100 GeV, 200 GeV and 300 GeV, in order to test the efficacy of such cut. As we are looking at the signal of single lepton with $\et$ hence $\et = p_T^\ell$. 
\item{} transverse mass ($m_T$) cut: we considered cuts of 200
  GeV, 300 GeV and 400 GeV. 
\end{itemize}

 The final summary table after implementing all the above mentioned cuts on
signal and backgrounds is given in Table \ref{table:1}. 
It is to be noted that the $p_T^\ell$ and $m_T$ distributions and cuts
are strongly correlated as they are related by eqn (\ref{eq:mt1}). 
The most efficient cut is $p_T^\ell > 300$ GeV, for which we listed the significance. 
In all the analysed benchmark 
points, a discovery is possible for ${\cal L} = 10$ fb$^{-1}$.
Harder cuts would reduce the signal and the overall statistics too much, therefore reducing the significance.

\subsection{ $\kappa_\ell \geq 0.5$} 

When $\kappa_\ell > 0.46$ the production cross-sections are typically
smaller due to the larger mass of the leptons, however the region $\kappa_\ell \geq 0.5$ can give 
interesting signatures at the LHC due to the presence of multiple leptons in the final state.
 The reason for this being that in
this region the T-odd leptons decay primarily via a charged $W_H$, which subsequently decays to $W$ and $A_H$.  
Following the decay chains in
eqns (\ref{eq:decay-chain}), we can get the following signature (triple
leptons): 
\beq 
pp \to \ell_H \nu_H \to \ell^\pm \ell^\mp \ell^\pm \et\ ,
\eeq
with a probability of about 1.5 \%.
This is a very interesting signature where the SM background is relatively
small and can be easily controlled. The SM backgrounds for trileptons
can be:
\begin{itemize}
\item{} $p p \to W^\pm Z$ where both $W$ and $Z$ decay leptonically,
$W^\pm \to \ell^\pm \bar{\nu}$ and $Z \to \ell \bar{\ell}$.
\item{} $ p p \to W^\pm W^\pm W^\mp$ where all the $W$'s decay
  leptonically, $W^\pm \to \ell^\pm \bar{\nu}$. 
\end{itemize}

In order to study the signal in this region of the LHT parameter space
with respect to the possible backgrounds at LHC we implemented the
following {\sl pre-selection cuts}:
\begin{itemize}
\item[(a)] jet veto: we apply a veto on events having a jet
  with $p_T > 30$ GeV within a rapidity of $|\eta| < 3$. 
\item[(b)] we demand that there are exactly three leptons, with
  only two of same charge, with $p_T > 20$ GeV
  and $|\eta| < 3$.  
\item[(c)] minimum $\et$ threshold of $30$ GeV.
\end{itemize}

In addition to the above mentioned {\sl pre-selection cuts}, to reduce the SM backgrounds we 
use the following secondary cuts:  
\begin{itemize}
\item{} invariant mass of the same sign leptons $|m_{\ell^\pm \ell^\mp} - m_Z| > 10$ GeV. This will reduce the
  backgrounds coming from leptons originating from a Z.
\item{} we demand that $|m_T(\ell \et) - m_W| > 15$. This will
  reduce the SM backgrounds coming from $W$'s.  
\item{} cut on $\et$ at 100 GeV: higher values for the cut would reduce too much the signal and the overall statistics. This cut could be
  helpful in reducing the backgrounds because, in the SM, $\et$ comes from
  neutrinos and hence could be relatively soft as compared to LHT models
  where $\et$ comes from heavy photons ($A_H$) can hence could be
  relatively hard.  
\end{itemize}

\begin{table}[htb]
\begin{center}
\begin{tabular}{| c | c | c || c | c |} 
\hline 
Process $\Rightarrow$ & LHT &  LHT & Background  & Background \\ 
Cuts $\Downarrow$ & $f = 500, \kappa_\ell = 0.5$ & $f = 500, \kappa_\ell = 0.55$ &
WZ & WWW    \\ \hline 
Preselection cuts                     &  39.1  &  102.9   &  14961.6  & 86.5 \\ 
$|m_{\ell^\pm \ell^\mp} - m_Z| >$ 10 GeV &  27.5  &  75.4    &  1032.2   & 65.9 \\
$|m_T(\ell \et) - m_W| >$ 15 GeV      &  26.1  &  72.2    &  609.1    & 56.25   \\ \hline
$\et > 100 $ GeV                      &  16.1  &  47.8    &  58.9    & 15.3  \\
$S/\sqrt{B}$                           &  1.9   &  5.5     &         &       \\   \hline 
\end{tabular}
\caption{\sl Efficiency of cuts on signal and background for
 the trilepton mode at LHC. The figures in the Table are number of
  events after each subsequent cut, assuming the integrated luminosity
  of LHC to be ${\cal L} = 300$ fb$^{-1}$. }   
\label{table:2}
\end{center}
\end{table}

At this point we would like to note that Datta {\sl et.al.}
\cite{Datta:2007xy} have also analysed the same signature (trilepton)
and have used similar kind of cuts to reduce the SM backgrounds. They
implemented the $\et > 100$ GeV cut before implementing a $m_T(\ell
\et)$ (transverse mass) cut and have indicated substantial reduction
in the backgrounds by imposing the $m_T$ cut. We disagree from their results
as $m_T$ and $\et$ are strongly correlated and hence one does not get
any substantial reduction in backgrounds by implementing a $\et$ cut
first and then imposing a $m_T$ cut. Accordingly we have implemented the $m_T$
cut first to reduce the backgrounds arising from
$W$-bosons. We have later imposed the $\et$ cut to further reduce the
backgrounds. 
We have estimated signal and major backgrounds for the above mentioned 
cuts in the trilepton mode: the results of our analysis are
summarised in table \ref{table:2}.  
We see that, even though the backgrounds are easily controlled, due to the small statistics, very large luminosities would be required at LHC 
in order to discover this channel.

\section{Pair production of charged T-odd leptons ($p p \to \ell_H
  \bar{\ell}_H$) \label{section:4}} 
  
\begin{figure}[htb]
\begin{center}
\epsfig{file=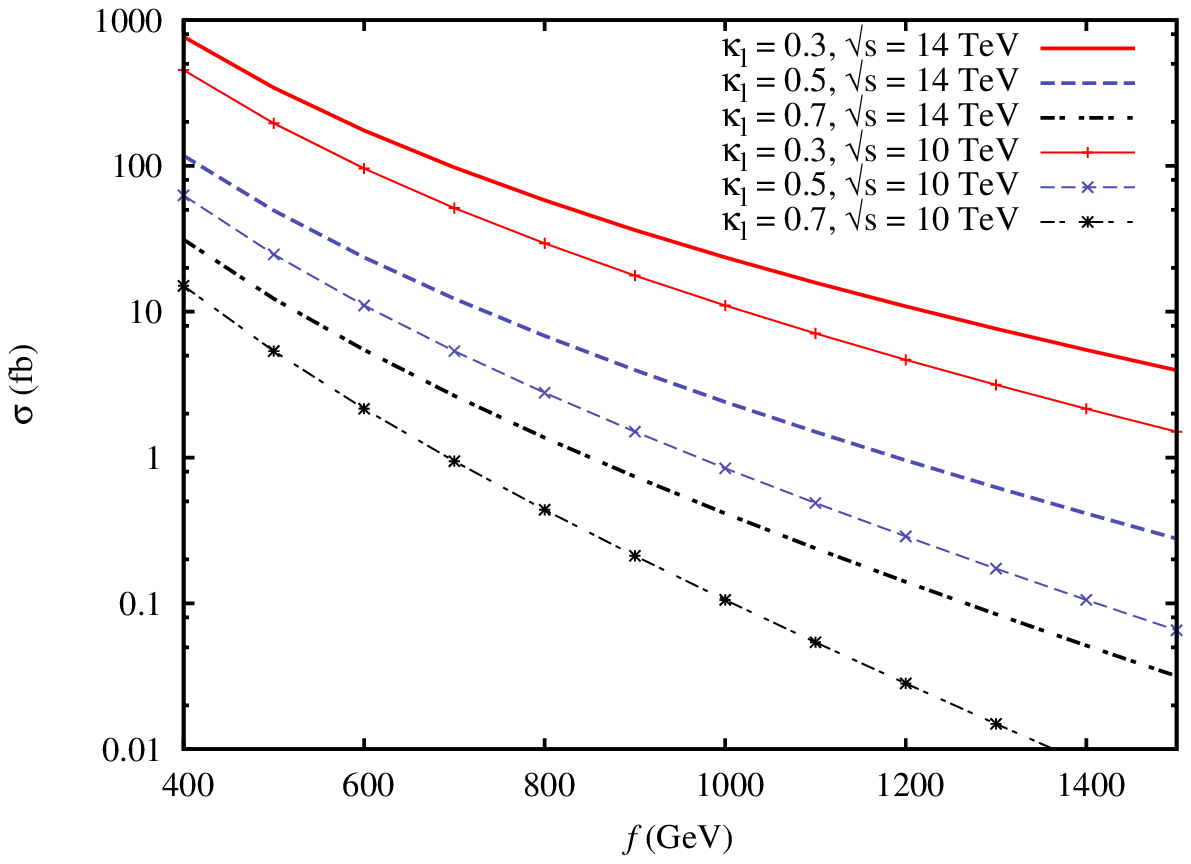,width=.8\textwidth}
\caption{\sl LHC production cross-sections for the process $\pptotll$.}
\label{fig:cs2}
\end{center}
\end{figure}

In Figure \ref{fig:cs2} we show the production cross-section for $p p \to \ell_H
\bar{\ell}_H$ as a function of the symmetry breaking scale $f$ for some
indicative values of $\kappa_\ell$ and for center of 
mass energy of 10 and 14 TeV. As can be
seen from Figures \ref{fig:cs1} and \ref{fig:cs2} the production
cross-section for pair of charged T-odd leptons is
relatively smaller than the process we have discussed in the previous
section. The reason for this being that in the channel discussed in
the previous section we also have to consider the charged conjugate
process.
The signature for $\pptotll$ in the $\kappa_\ell < 0.46$ region of the
parameter space is two opposite sign leptons and $\et$. 
As analysed in Ref. \cite{Godbole:2003it}, in this case it is
relatively easier to suppress backgrounds
as compared to single charged lepton with $\et$.
For $\kappa_l > 0.46$, the decay chains in eqn \ref{eq:decay-chain}
would also give rise to the same signature.
 However, the effective cross-section in this case would be too
small considering the branching ratio of $\ell_H \to W_H \nu$ and, after 
the decay $W_H \to W A_H$, of the leptonic decay of the $W$.
Therefore in our analysis we will restrict
ourselves to the charged pair channel with $\kappa_\ell < 0.46$.

The main backgrounds to opposite sign lepton pair and $\et$ signal
can come from:
\begin{itemize} 
\item{} $pp \to W^+ W^- \to
  \ell^\pm \ell^\mp \et$  where the
  charged gauge boson decays leptonically.
\item{} $p p \to Z  Z  \to
  \ell^\pm \ell^\mp \et$ where one of the $Z$'s decays leptonically and the other invisibly.
\end{itemize}
We propose to use following {\sl pre-selection cuts}: 
\begin{itemize}
\item[(a)] jet veto: veto events having jets with $p_T > 30$ GeV and
  $|\eta| < 3$. 
\item[(b)] exactly two leptons of opposite charge. The leptons must be
  visible in the detector so we require them to have $p_T > 10 $
  GeV and $|\eta| < 3$. 
\item[(c)] missing energy threshold $\et > 30$ GeV. 
\end{itemize}

\begin{figure}[htb]
\begin{center}
\epsfig{file=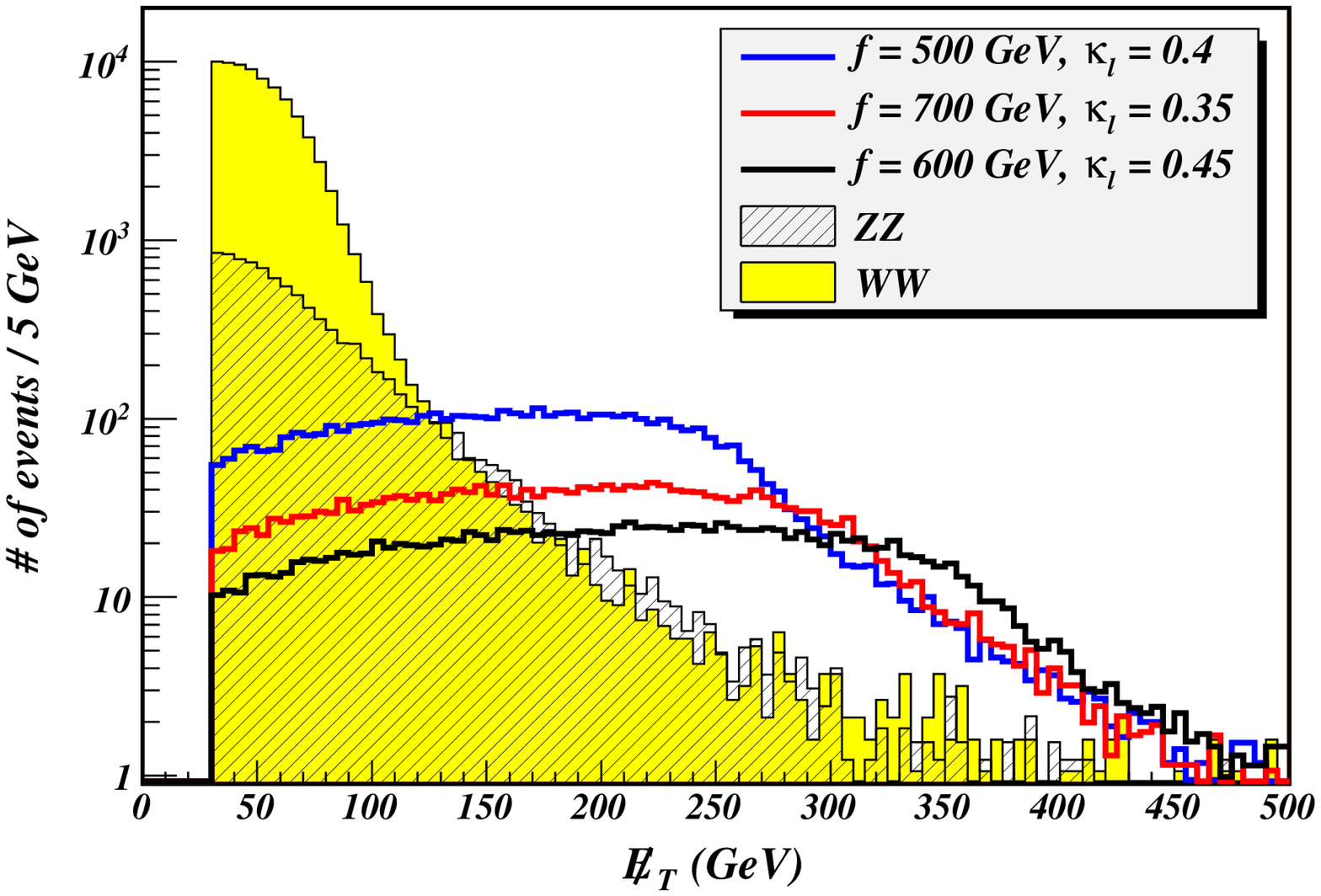,width=.5\textwidth} 
\hskip -0.5cm
\epsfig{file=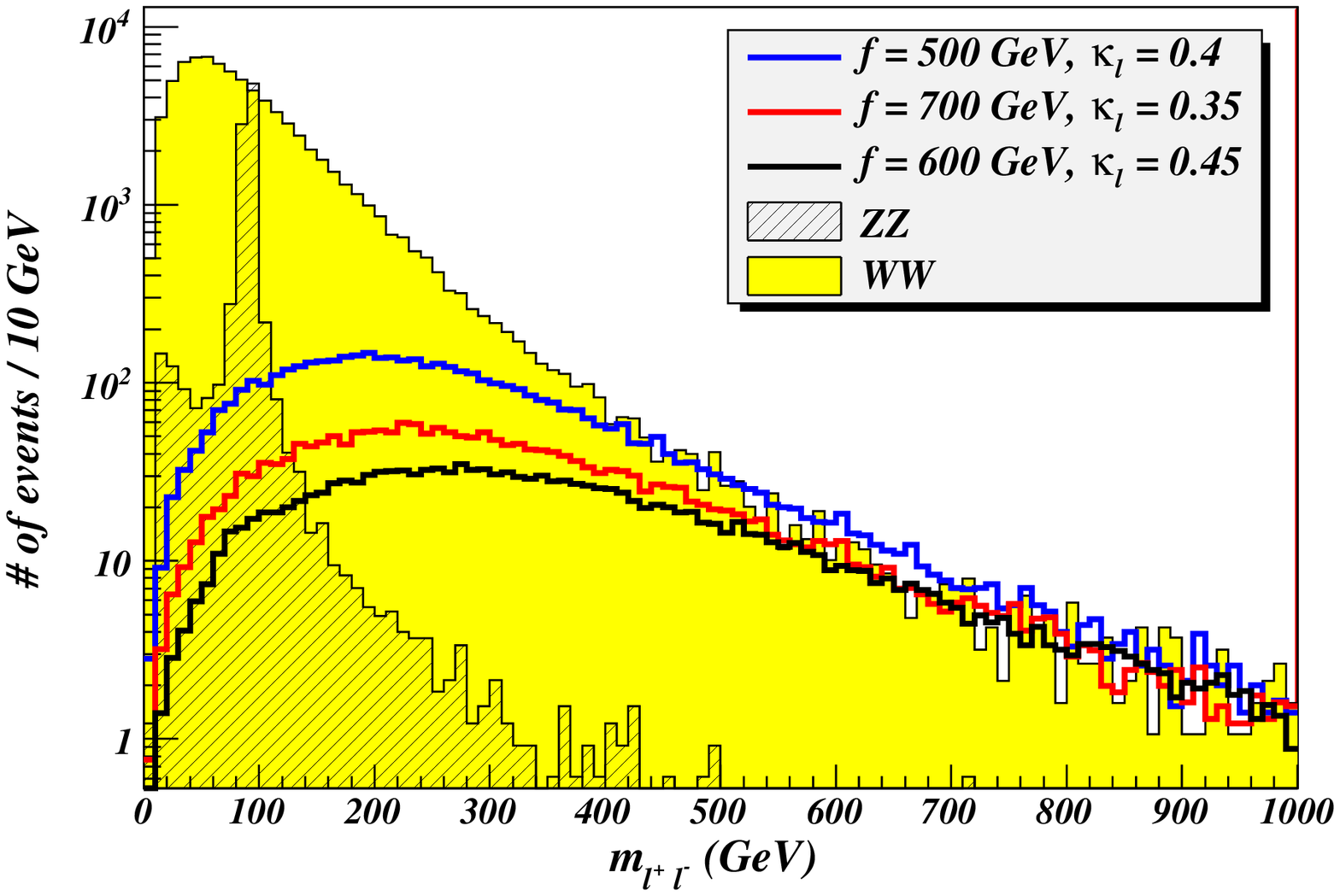,width=.5\textwidth}
\caption{\sl $\et$ distribution (left) and lepton invariant mass
($m_{\ell \bar{\ell}}$) distribution (right) for the process
$pp \to \ell_H \bar{\ell}_H \to \ell \bar{\ell} + \et$ with LHC luminosity of $100$ fb$^{-1}$, including backgrounds. The model
parameters are given in the legends.}
\label{fig:tlep2}
\epsfig{file=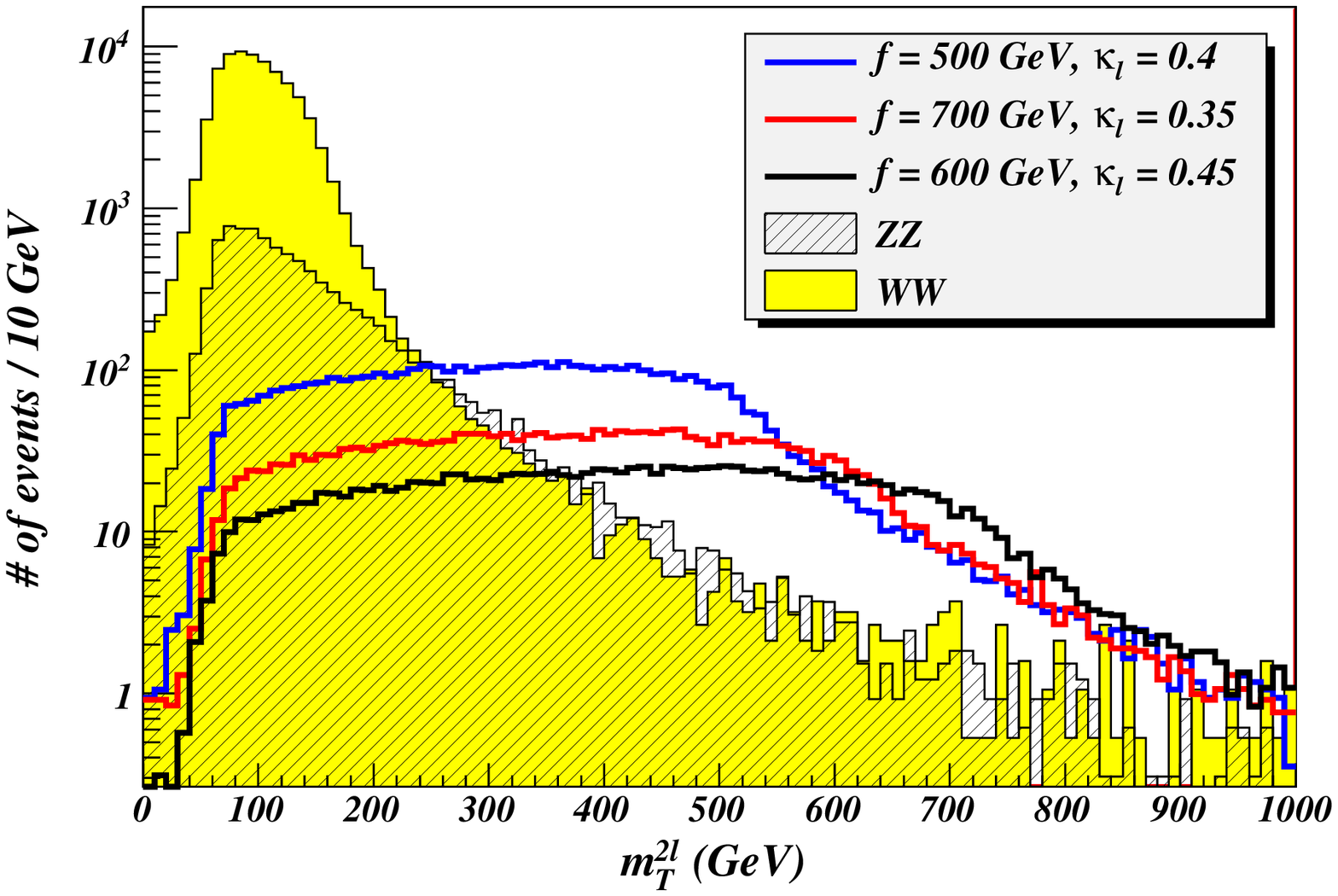,width=.5\textwidth}
\caption{\sl $m_T^{2\ell}$ distribution for the process 
$pp \to \ell_H \bar{\ell}_H \to \ell \bar{\ell} + \et$ with LHC
luminosity of ${\cal L} = 100$ fb$^{-1}$, including backgrounds. }  
\label{fig:tlep3}
\end{center}
\end{figure}

The results of the signal and background events after imposing {\sl
  pre-selection} cuts are given in Table \ref{table:3}. As can be seen
from the table the background events surviving the {\sl pre-selection
cuts} are orders magnitude greater than the signal events. More cuts are therefore necessary in order to
improve the signal as compared to backgrounds. To device the {\sl secondary cuts} we
have plotted the $\et$ and dilepton invariant mass ($m_{\ell \ell}$)
distributions for some signal point and backgrounds in Figure
\ref{fig:tlep2}.  
We also define the two dilepton transverse invariant
mass as \cite{Godbole:2003it}:
\beq
 m_T^{2\ell} = \sqrt{2 p_T^{2 \ell} \et (1 - cos \phi) }\ ,
 \eeq
where $p_T^{2\ell}$ is the sum of the transverse momenta of the two
leptons. The $m_T^{2\ell}$ distribution is shown in Figure
\ref{fig:tlep3}. By observing these distributions we propose to use
the  following secondary cuts:
\begin{itemize}
\item{} $\et >$ 100 GeV. 
\item{} the invariant mass of the lepton pair is away from $m_Z$, $|m_{\ell \bar{\ell}} - m_Z | > 10$
  GeV. This
  reduces the backgrounds where a lepton pair originates from
  Z-decay. 
  \item{} transverse mass cut, $m_T^{2 \ell} > 200$ GeV. 
\end{itemize}
The summary of our results for some particular sets of input parameters
$(f,\kappa_\ell)$ is shown in Table \ref{table:3}. 
The $\et$ and $m_{\ell \bar{\ell}}$ cuts are very effective in reducing the $WW$ and $ZZ$ cuts without affecting the signal.
Therefore, this channel offers a powerful discovery potential even at integrated luminosity as low as 1 fb$^{-1}$.

\begin{table}
\begin{center}
\begin{tabular}{| c | c | c || c | c | c | c | c |} 
\hline 
Model parameters $\Rightarrow$ & SM  & SM  & $f = 500$  & $f=600$ & $f = 700$ & $f
= 600$ & $f = 700$ \\ 
Cuts $\Downarrow$  &  WW & ZZ  & $\kappa_\ell = 0.4$ & $\kappa_\ell =
0.35$ & $\kappa_\ell = 0.3$ & $\kappa_\ell = 0.45$ & $\kappa_\ell =
0.35$ \\ \hline 
$\sigma$ (fb)      &  &  &  117.6   & 97.5    & 97.58   & 36.27  &
53.48   \\ \hline 
Pre-selection cuts & 7.67 $\times 10^4$ & 9316.8 & 4738 & 3918.9 & 3993.4 & 1449.5 & 2142.8  \\ \hline
$\et > $ 100 GeV   & 1994.1 &  1672.8 &  3669 & 3071.7  & 3065.5 & 1247.6 & 1767.3  \\
$|m_{\ell \bar{\ell}} - m_Z| > 10$ GeV & 1814.1 & 233.7 & 3496.7 & 2942.1 & 2869.3 & 1216.2 & 1710.7   \\ 

$m_T^{2 \ell} > 200 $ GeV & 1419.2 & 210.4 & 3433.6 & 2890.8 & 2869.3 & 1202.9 & 1687.8 \\ \hline
$S/\sqrt{B}$(1 fb$^{-1}$) &   &   & 8.5 & 7.1 & 7.1 & 3 & 4.2 \\
$S/\sqrt{B}$(3 fb$^{-1}$) &   &   & 14.9 & 12.4 & 12.4 & 5.2 & 7.3 \\
$S/\sqrt{B}$(10 fb$^{-1}$) &   &   & 26.9 & 22.6 & 22.5 & 9.4 & 13.2 \\
\hline  
\end{tabular}
\caption{\sl Results of the simulations in the channel $\ell^+
  \ell^- \et$. The above numbers indicate the number of
  events after imposing of sequential selection cuts as defined in
  the text for an LHC luminosity ${\cal L} = 100$ fb$^{-1}$. 
 We have also given the significance $S/\sqrt{B}$ after the cuts for various integrated luminosities.} 
\label{table:3}
\end{center}
\end{table}

\section{Pair production of neutral T-odd leptons ($p p \to \nu_H
  \bar{\nu}_H$) \label{section:5}}

\begin{figure}[htb]
\begin{center}
\epsfig{file=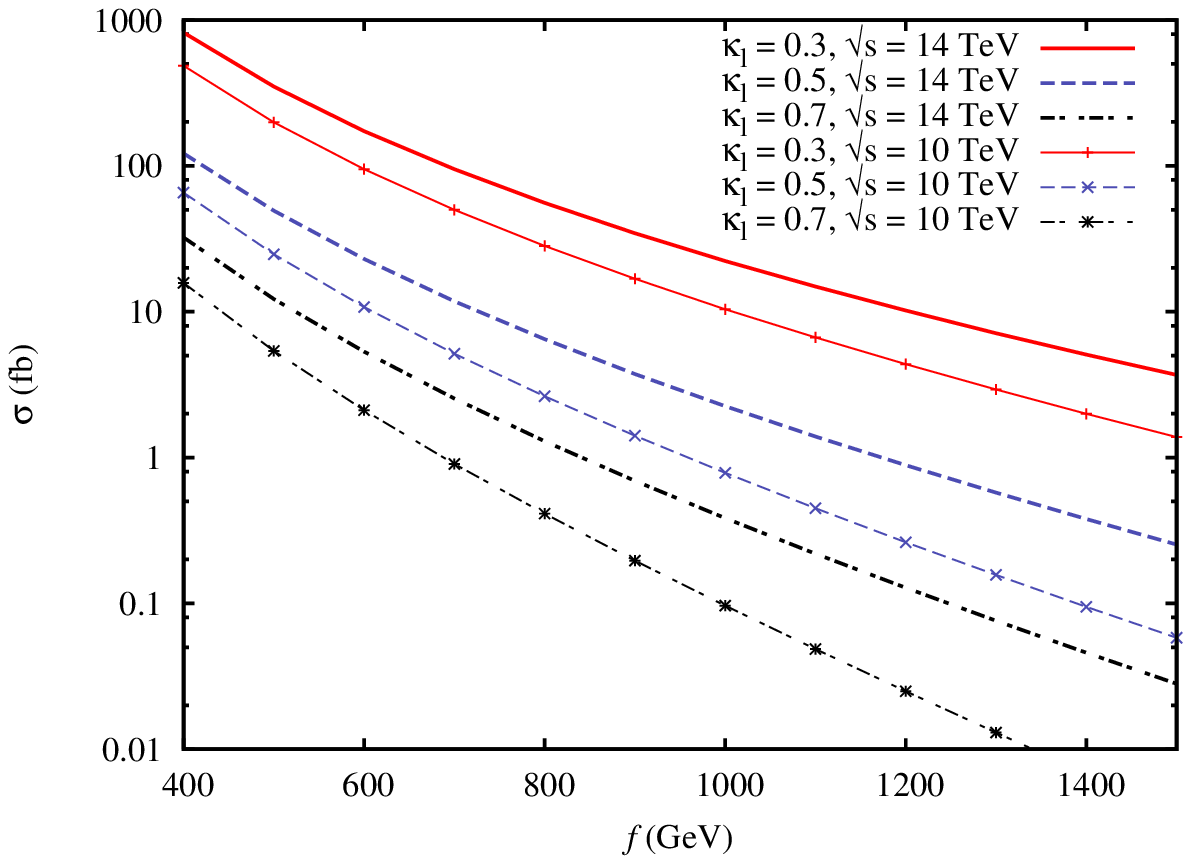,width=0.63\textwidth}
\caption{\sl LHC production cross-sections for the process $p p \to
  \nu_H \bar{\nu}_H$ for the options of centre of mass energy of 14
  TeV and 10 TeV, and for different benchmark points as a function of
  $f$.}   \label{fig:csnu}
\end{center}
\end{figure}

For completeness, we discuss here the pair production of heavy neutral T-odd leptons. 
The production cross-sections $p p \to \nu_H
\bar{\nu}_H$, shown in Figure~\ref{fig:csnu}, are of the same order as the ones in the previous section.
However, for $\kappa_\ell < 0.46$ the neutral leptons decay invisibly into neutrino and heavy photon, 
thus not leaving any signatures. 
On the other hand, for $\kappa_l > 0.46$, the decay chain in eqn  \ref{eq:decay-chain} gives rise to 
a very clean four lepton channel.
The rate of such events, however, is very small due to the smallness of the cross-sections and 
the suppression of the branching ratios.

\noindent In the SM, four leptons can arise from:
\begin{itemize}
\item{} $p p \to W^\pm W^\mp Z$ with the gauges bosons decaying leptonically. 
\item{} $p p \to Z Z Z$ with two Z's decaying via $Z \to \ell^\pm
  \ell^\pm$ and third Z decaying via $Z \to \nu \bar{\nu}$.
\item{} $p p \to Z Z$ with each of Z decaying leptonically. 
\end{itemize}

\noindent The {\sl pre-selection} cuts we used are:
\begin{itemize}
\item[(a)] exactly four charged leptons, two each of same charge with
  $p_T^\ell > 15$ GeV and $|\eta| < 3$. 
\item[(b)] jet veto: no jet with $p_T > 30$ GeV and $|\eta| < 3$ in
  the event. 
\end{itemize}

The {\sl secondary} cuts imposed to reduce backgrounds are : 
\begin{itemize}
\item{} invariant mass of the same sign leptons $|m_{\ell^\pm
    \ell^\mp} - m_Z| > 10$ GeV. This will reduce the backgrounds
  coming from leptons originating from a Z. 
\item{} we demand that $|m_T(\ell \et) - m_W| > 15$. This will reduce
  the SM backgrounds coming from $W$'s.   
\item{} cut on $\et$ at 100 GeV: higher values for the cut would
  reduce too much the signal and the overall statistics. 
  \end{itemize}

\begin{table}[htb]
\begin{center}
\begin{tabular}{|c | c | c | c | c | c |}\hline 
Parameter set $\Rightarrow$ &   $f = 500$   &  $f = 500$  & SM & SM & SM  \\  
Cuts $\Downarrow$   &      $\kappa = 0.5$ & $\kappa = 0.55$ & ZZW &
ZZZ & ZZ \\ \hline 
$\sigma$ (fb) & 49.3 & 33.5  &   &  &   \\ \hline 
Pre-selection & 2.1 & 29.6 & 14.9 & 2.7 & 44.9 \\ 
$|m_{\ell^+ \ell^-} - m_Z| > 10$ GeV & 1.5 & 19.7 & 6.9 & 0.4 & 9.8 \\ 
$|m_T(\ell \et) - m_W| > 15$ GeV & 1.3 & 19 &  6.1 & 0.3 & 8.4 \\ \hline
$\et > 100$ GeV & 0.7 & 12.1 & 2.4 & 0.2 & 5.6 \\ 
$S/\sqrt{B}$ &  0.2 & 4.2 & & & \\ \hline 
\end{tabular}
\caption{\sl Number of signal and background events for ${\cal L} =
  300$ fb$^{-1}$. } \label{table:4}
\end{center}
\end{table}

The results of our simulations are summarized in Table \ref{table:4}. 
As in the trilepton case, the small statistics requires very large 
integrated luminosities in order to observe this channel, even though the backgrounds are easily controlled.

\section{Conclusions \label{section:6}}

In this work we have discussed the phenomenology of T-parity odd heavy leptons at the LHC in the Littlest Higgs 
model as a sample of the corresponding phenomenology for typical Little Higgs models with 
T-parity. 
This type of models predicts a set of new T-odd fermions in addition to the heavy gauge bosons
of the Little Higgs model. 
We have studied T-odd charged lepton single and pair production at the LHC and their purely leptonic decays.
Production of a pair of heavy neutrinos, when giving visible leptonic signatures, have a too feeble rate to be detected.
Those channels are very clean at the LHC due to the absence of jets which can rid of most of the QCD background.
In the single charged channel, the production cross-sections at LHC are large and can be more than a picobarn in part of the 
$(f,\kappa_\ell)$ parameter space. For $\kappa_\ell < 0.46$ the heavy leptons decay only to the heavy photon $A_H$ 
and the corresponding standard model lepton. This yields the single lepton signature $\ell^\pm \et$, which can be discovered at LHC over the 
background with an integrated luminosity of around 10 fb$^{-1}$ both for the 10 TeV and the 14 TeV centre of mass energy options.
When $\kappa_\ell > 0.46$ the production cross-sections are typically small due to the larger mass of the heavy leptons, and decay modes involving the heavy $W_H$ and $Z_H$ bosons open up.
For $\kappa_l > 0.5$ this gives rise to a trilepton signature, however it is suppressed by small branching ration and therefore a high integrated luminosity of 300 fb$^{-1}$ is necessary for the observation of such channel. 

The process $\pptotll$ has a smaller production cross-section which is typically in the $100$ femtobarn region, however it is easier to detect over the
backgrounds due to the presence of two opposite charge leptons in the final state. 
For $\kappa_l < 0.46$, the dilepton plus missing energy ($\ell^+  \ell^- \et$) signal can be easily detected at LHC over the background, and an integrated luminosity of 1 to 3 fb$^{-1}$ is sufficient for the discovery.

In summary, for small values of the Yukawa coupling $\kappa_l$, the T-odd heavy leptons of T-parity little Higgs models are easily detectable at the LHC due to the purely leptonic signatures.
This provides is a new handle to test this type of models
with a few spectacular channels that can be studied even in the early stage of the LHC running, with clear and visible 
signatures over the background.
For larger Yukawa couplings, the purely leptonic signal is suppressed by branching ratios, therefore requiring very large integrated luminosities for the detection.
In this case, semi-leptonic decays may be interesting, however we leave their study to future work.

\section*{Acknowledgements}
We would like to thank Satyaki Bhattacharya, Debajyoti Choudhury,
and Sudhir K. Gupta for useful discussions/communications.  
The work of SRC was supported by Ramanna fellowship of Department of
Science \& Technology (DST), India.



\begin{thebibliography}{99}


\bibitem{littlehierarchy}
  R.~Barbieri and A.~Strumia,
  arXiv:hep-ph/0007265.

\bibitem{lhrev}
  M.~Schmaltz and D.~Tucker-Smith, 
  Ann.\ Rev.\ Nucl.\ Part.\ Sci.\  {\bf 55}, 229 (2005)
  [arXiv:hep-ph/0502182],
 M.~Perelstein,
  Prog.\ Part.\ Nucl.\ Phys.\  {\bf 58}, 247 (2007)
  [arXiv:hep-ph/0512128].

\bibitem{precisionEW}
  C.~Csaki, J.~Hubisz, G.~D.~Kribs, P.~Meade and J.~Terning,
  Phys.\ Rev.\  D {\bf 67}, 115002 (2003)
  [arXiv:hep-ph/0211124]
and
  Phys.\ Rev.\  D {\bf 68}, 035009 (2003)
  [arXiv:hep-ph/0303236];
  T.~Gregoire, D.~Tucker-Smith and J.~G.~Wacker,
  Phys.\ Rev.\  D {\bf 69}, 115008 (2004)
  [arXiv:hep-ph/0305275];
  R.~Casalbuoni, A.~Deandrea and M.~Oertel,
  JHEP {\bf 0402}, 032 (2004)
  [arXiv:hep-ph/0311038];
  W.~Kilian and J.~Reuter,
  Phys.\ Rev.\ D {\bf 70}, 015004 (2004)
  [arXiv:hep-ph/0311095];
  G.~Marandella, C.~Shappacher and A.~Strumia,
  Phys.\ Rev.\  D {\bf 72}, 035014 (2005)
  [arXiv:hep-ph/0502096];
  Z.~Han and W.~Skiba,
  Phys.\ Rev.\  D {\bf 72}, 035005 (2005)
  [arXiv:hep-ph/0506206].
  
  
\bibitem{tpar}
  H.~C.~Cheng and I.~Low,
  JHEP {\bf 0309}, 051 (2003)
  [arXiv:hep-ph/0308199],
H.~C.~Cheng and I.~Low,
  JHEP {\bf 0408}, 061 (2004)
  [arXiv:hep-ph/0405243],
H.~C.~Cheng, I.~Low and L.~T.~Wang,
  Phys.\ Rev.\  D {\bf 74}, 055001 (2006)
  [arXiv:hep-ph/0510225];
%
  I.~Low,
  JHEP {\bf 0410}, 067 (2004)
  [arXiv:hep-ph/0409025].


\bibitem{tparprec}
J.~Hubisz, P.~Meade, A.~Noble and M.~Perelstein,
  JHEP {\bf 0601}, 135 (2006)
  [arXiv:hep-ph/0506042].


\bibitem{tparDM}
A.~Birkedal, A.~Noble, M.~Perelstein and A.~Spray,
Phys.\ Rev.\  D {\bf 74}, 035002 (2006)
  [arXiv:hep-ph/0603077].





\bibitem{Hubisz:2004ft}
  J.~Hubisz and P.~Meade,
  Phys.\ Rev.\  D {\bf 71}, 035016 (2005)
  [arXiv:hep-ph/0411264];
%
  J.~Hubisz, P.~Meade, A.~Noble and M.~Perelstein,
  JHEP {\bf 0601}, 135 (2006)
  [arXiv:hep-ph/0506042].


\bibitem{Goto:2008fj}
  T.~Goto, Y.~Okada and Y.~Yamamoto,
  Phys.\ Lett.\  B {\bf 670}, 378 (2009)
  [arXiv:0809.4753 [hep-ph]]; 
%
  F.~del Aguila, J.~I.~Illana and M.~D.~Jenkins,
  JHEP {\bf 0901}, 080 (2009)
  [arXiv:0811.2891 [hep-ph]];
%
  M.~Blanke, A.~J.~Buras, B.~Duling, S.~Recksiegel and C.~Tarantino,
  arXiv:0906.5454 [hep-ph].


\bibitem{Choudhury:2006sq}
  S.~R.~Choudhury, A.~S.~Cornell, A.~Deandrea, N.~Gaur and A.~Goyal,
  Phys.\ Rev.\  D {\bf 75} (2007) 055011
  [arXiv:hep-ph/0612327];
%
  M.~Blanke, A.~J.~Buras, B.~Duling, A.~Poschenrieder and C.~Tarantino,
  JHEP {\bf 0705} (2007) 013
  [arXiv:hep-ph/0702136];
%
  N.~Gaur,
  AIP Conf.\ Proc.\  {\bf 981}, 357 (2008)
  [arXiv:0710.3998 [hep-ph]].


\bibitem{Pukhov:2004ca}
  A.~Pukhov,
  arXiv:hep-ph/0412191.

\bibitem{Belyaev:2006jh}
  A.~Belyaev, C.~R.~Chen, K.~Tobe and C.~P.~Yuan,
  Phys.\ Rev.\  D {\bf 74}, 115020 (2006)
  [arXiv:hep-ph/0609179].

\bibitem{Sjostrand:2006za}
  T.~Sjostrand, S.~Mrenna and P.~Skands,
  JHEP {\bf 0605}, 026 (2006)
  [arXiv:hep-ph/0603175].


\bibitem{Alwall:2007mw}
  J.~Alwall {\it et al.},
  arXiv:0712.3311 [hep-ph]; 
%
  E.~Boos {\it et al.},
  arXiv:hep-ph/0109068 ; 
%
  J.~Alwall {\it et al.},
  Comput.\ Phys.\ Commun.\  {\bf 176}, 300 (2007)
  [arXiv:hep-ph/0609017].


\bibitem{atlfast}
 E. Richter-Was {\sl et.al.} , {\it ATLFAST 2.2: A fast simulation package for
 ATLAS}, ATL-PHYS-98-131. 


\bibitem{Maltoni:2002qb}
  F.~Maltoni and T.~Stelzer,
  JHEP {\bf 0302}, 027 (2003)
  [arXiv:hep-ph/0208156].


\bibitem{Cao:2007pv}
  Q.~H.~Cao and C.~R.~Chen,
  Phys.\ Rev.\  D {\bf 76}, 075007 (2007)
  [arXiv:0707.0877 [hep-ph]].


\bibitem{Datta:2007xy}
  A.~Datta, P.~Dey, S.~K.~Gupta, B.~Mukhopadhyaya and A.~Nyffeler,
  Phys.\ Lett.\  B {\bf 659}, 308 (2008)
  [arXiv:0708.1912 [hep-ph]].


\bibitem{Godbole:2003it}
  R.~M.~Godbole, M.~Guchait, K.~Mazumdar, S.~Moretti and D.~P.~Roy,
  Phys.\ Lett.\  B {\bf 571}, 184 (2003)
  [arXiv:hep-ph/0304137].

\bibitem{Cao:2004tu}
  Q.~H.~Cao, S.~Gopalakrishna and C.~P.~Yuan,
  Phys.\ Rev.\  D {\bf 70}, 075020 (2004)
  [arXiv:hep-ph/0405220].





\end{thebibliography}
\end{document}